\documentclass[12pt,epsf]{article}
\usepackage{graphicx,amsmath,amssymb}
\usepackage{epstopdf}
\begin{document}

\def\a{\alpha}
\def\b{\beta}
\def\c{\varepsilon}
\def\d{\delta}
\def\e{\epsilon}
\def\f{\phi}
\def\g{\gamma}
\def\h{\theta}
\def\k{\kappa}
\def\l{\lambda}
\def\m{\mu}
\def\n{\nu}
\def\p{\psi}
\def\q{\partial}
\def\r{\rho}
\def\s{\sigma}
\def\t{\tau}
\def\u{\upsilon}
\def\v{\varphi}
\def\w{\omega}
\def\x{\xi}
\def\y{\eta}
\def\z{\zeta}
\def\D{\Delta}
\def\G{\Gamma}
\def\H{\Theta}
\def\L{\Lambda}
\def\F{\Phi}
\def\P{\Psi}
\def\S{\Sigma}

\def\o{\over}
\def\beq{\begin{eqnarray}}
\def\eeq{\end{eqnarray}}
\newcommand{\gsim}{ \mathop{}_{\textstyle \sim}^{\textstyle >} }
\newcommand{\lsim}{ \mathop{}_{\textstyle \sim}^{\textstyle <} }
\newcommand{\vev}[1]{ \left\langle {#1} \right\rangle }
\newcommand{\bra}[1]{ \langle {#1} | }
\newcommand{\ket}[1]{ | {#1} \rangle }
\newcommand{\EV}{ {\rm eV} }
\newcommand{\KEV}{ {\rm keV} }
\newcommand{\MEV}{ {\rm MeV} }
\newcommand{\GEV}{ {\rm GeV} }
\newcommand{\TEV}{ {\rm TeV} }
\def\diag{\mathop{\rm diag}\nolimits}
\def\Spin{\mathop{\rm Spin}}
\def\SO{\mathop{\rm SO}}
\def\O{\mathop{\rm O}}
\def\SU{\mathop{\rm SU}}
\def\U{\mathop{\rm U}}
\def\Sp{\mathop{\rm Sp}}
\def\SL{\mathop{\rm SL}}
\def\tr{\mathop{\rm tr}}

\def\IJMP{Int.~J.~Mod.~Phys. }
\def\MPL{Mod.~Phys.~Lett. }
\def\NP{Nucl.~Phys. }
\def\PL{Phys.~Lett. }
\def\PR{Phys.~Rev. }
\def\PRL{Phys.~Rev.~Lett. }
\def\PTP{Prog.~Theor.~Phys. }
\def\ZP{Z.~Phys. }

\newcommand{\bea}{\begin{eqnarray}}   
\newcommand{\eea}{\end{eqnarray}}
\newcommand{\bear}{\begin{array}}  
\newcommand {\eear}{\end{array}}
\newcommand{\bef}{\begin{figure}}  
\newcommand {\eef}{\end{figure}}
\newcommand{\bec}{\begin{center}}  
\newcommand {\eec}{\end{center}}
\newcommand{\non}{\nonumber}  
\newcommand {\eqn}[1]{\beq {#1}\eeq}
\newcommand{\la}{\left\langle}  
\newcommand{\ra}{\right\rangle}
\newcommand{\ds}{\displaystyle}
\def\SEC#1{Sec.~\ref{#1}}
\def\FIG#1{Fig.~\ref{#1}}
\def\EQ#1{Eq.~(\ref{#1})}
\def\EQS#1{Eqs.~(\ref{#1})}
\def\GEV#1{10^{#1}{\rm\,GeV}}
\def\MEV#1{10^{#1}{\rm\,MeV}}
\def\KEV#1{10^{#1}{\rm\,keV}}
\def\lrf#1#2{ \left(\frac{#1}{#2}\right)}
\def\lrfp#1#2#3{ \left(\frac{#1}{#2} \right)^{#3}}


\baselineskip 0.7cm

\begin{titlepage}

\begin{flushright}
TU-952\\
IPMU14-0010\\
\end{flushright}

\vskip 1.35cm
\begin{center}
{\large \bf 
Multi-Natural Inflation
}
\vskip 1.2cm
Michael Czerny$^{a}$\footnote{email: mczerny@tuhep.phys.tohoku.ac.jp} and
Fuminobu Takahashi$^{a,b}\footnote{email: fumi@tuhep.phys.tohoku.ac.jp}$

\vskip 0.4cm
{\it $^a$Department of Physics, Tohoku University, Sendai 980-8578, Japan}\\
{\it $^b$Kavli Institute for the Physics and Mathematics of the Universe (WPI), TODIAS, University of Tokyo, Kashiwa 277-8583, Japan}

\vskip 1.5cm

\abstract{
We propose a multi-natural inflation model in which the single-field inflaton potential consists of two  or more
sinusoidal potentials that are comparable in size, but have different periodicity with a possible non-zero relative phase.  
The model is versatile enough to realize both large-field and small-field inflation. 
We show that,  in a model with two sinusoidal potentials,
the predicted values of the spectral index and tensor-to-scalar ratio lie within the $1\sigma$ region of the Planck data.
In particular,  
there is no lower bound on the decay constants in contrast to the original natural inflation.
We also show that,
 in a certain limit, multi-natural inflation can be approximated by a hilltop quartic inflation model.
}
\end{center}
\end{titlepage}

\setcounter{page}{2}

\section{Introduction}

The standard  $\Lambda$CDM cosmology has been strongly confirmed by recent
Planck observations~\cite{Ade:2013rta}. In particular, the observed data is consistent with
almost scale-invariant, adiabatic and Gaussian primordial density perturbations. 
This strongly suggests that our Universe experienced 
the inflationary expansion described by a simple (effectively) single-field 
inflation~\cite{Guth:1980zm,Linde:1981mu}. 

Among many inflation models so far, there is an interesting class of models called
large-field inflation in which  sizeable tensor perturbations are generated. 
One of the large-field inflation models is chaotic inflation~\cite{Linde:1983gd}, and by now
there are many concrete realizations of the chaotic inflation in supergravity~\cite{Kawasaki:2000yn,Kawasaki:2000ws,
Kallosh:2010ug,Kallosh:2010xz,Takahashi:2010ky,Nakayama:2010kt,Harigaya:2012pg} 
and superstring theory~\cite{Silverstein:2008sg,McAllister:2008hb} (see also e.g. Refs.~\cite{Croon:2013ana,
Nakayama:2013jka,Ellis:2013xoa,Kallosh:2013pby} for the development after the Planck results).

The tensor-to-scalar ratio $r$ as well as the spectral index $n_s$ are tightly constrained
by the Planck data  combined with other CMB  observations as~\cite{Ade:2013rta}
\begin{align}
n_s &= 0.9600 \pm 0.0071, \\
r &< 0.11 ~~~(95\% {\rm CL}).
\end{align}
While $n_s$ is determined by the shape of the inflaton potential, $r$ is determined by
the inflation energy scale, as long as the slow-roll inflation is assumed. Then
the inflation scale is related to $r$ as
\beq
H_{\rm inf} \simeq 8.5 \times 10^{13} \,{\rm GeV} \lrfp{r}{0.11}{\frac{1}{2}},
\eeq
where $H_{\rm inf}$ denotes the Hubble parameter during inflation, and the on-going and 
planned CMB observations will be able to probe $r \gtrsim 10^{-3}$.

In a single-field inflation model with a canonical kinetic term, there is a relation between $r$ and 
the field excursion of the inflaton~\cite{Lyth:1996im}. In particular, 
the inflaton field excursion exceeds the Planck scale if $r \gtrsim 0.01$, which places a  strict requirement 
on inflation model building to have good control over inflaton field values greater than the Planck scale.
One possibility is to impose an approximate shift symmetry on the inflaton so as to keep the inflaton 
potential sufficiently flat at super-Planckian values. The simplest realization is natural inflation~\cite{Freese:1990ni,Adams:1993ni},
in which the inflaton is a (pseudo) Nambu-Goldstone (NG) boson, and its potential takes the following form, 
\begin{equation}
  V(\f) = \Lambda^4\left[1 - \cos\left(\f/f\right)\right],
  \label{ni}
\end{equation}
where the sinusoidal potential arises from some non-perturbative effects which explicitly break
the shift symmetry. 
The predicted values of $n_s$ and $r$ are consistent with the Planck data for $f \gtrsim 5 M_p$~\cite{Ade:2013rta},
where $M_p \simeq 2.4 \times \GEV{18}$ is the reduced Planck mass.

In this letter we consider an extension of natural inflation by adding another sinusoidal potential
which modifies the inflaton dynamics at large field values, leading to different predictions of $n_s$ and $r$.\footnote{
Increasing the number of parameters is not favored from the Bayesian point of view.
}
This is possible if there are multiple sources for the explicit breaking of the shift symmetry of the inflaton~\cite{Takahashi:2013tj}.
We shall give concrete examples later.\footnote{
Although our model is a simple toy model at this stage, the implementation in string-inspired supergravity~\cite{Czerny:2014xja} 
as well as its implications for a large running spectral index~\cite{Czerny:2014wua} were studied after the submission of this paper.
}
Our model is versatile enough to realize both large-field
and small-field inflation.
We will show that  the predicted values of $n_s$ and $r$ are consistent with the Planck data
for a wide range of the decay constant.   In particular, the sub-Planckian decay constant $f  \ll M_p$
is allowed by the Planck data at $2\sigma$; the model is approximated by a hilltop quartic inflation model~\cite{Linde:1981mu} in this limit.

Lastly let us briefly mention related works in the past. 
In the original natural inflation, the required decay constant $f\gtrsim 5 M_p$ may be beyond the range of validity
of an effective field theoretic description.  One solution is to consider multiple axions (or NG bosons); in Ref.~\cite{Kim:2004rp},
it was pointed out that the effective large decay constant can be realized, leading to the (effectively) single-field
natural inflation (\ref{ni}). 
See also Refs.~\cite{Mohanty:2008ab,Germani:2010hd} for other ways to relax the bound.
 Using pseudo NG bosons, multi-field inflation models such as hybrid inflation were 
proposed in Refs.~\cite{Kaplan:2003aj}. In string theory, there are racetrack inflation~\cite{BlancoPillado:2004ns}, 
$N$-flation~\cite{Dimopoulos:2005ac}, and axion monodromy~\cite{Silverstein:2008sg,McAllister:2008hb},
in which the axions play the role of the inflaton; the first two models are multi-field inflation models, and the last one is
equivalent to a linear-term chaotic inflation.  Our model is a single-field inflation model based on multiple sinusoidal functions, 
and the inflaton potential as well as the predicted $n_s$ and $r$ are different 
from those of the natural inflation and the above models. Later we will briefly discuss a possible UV completion of our model.

\section{Multi-natural inflation}
\label{sec:2}

\subsection{Natural inflation}
Here let us summarize the results of natural inflation.
Natural inflation arises from a 
broken global symmetry in order to generate a very flat potential necessary for inflation~\cite{Freese:1990ni,Adams:1993ni}.
The  inflaton potential has the form
\begin{equation}
  V(\f) = \Lambda^4\left[1 - \cos(\f/f)\right].
\end{equation}
In a standard slow roll analysis there are two sufficient conditions for inflation, given by
\begin{equation}
  \varepsilon(\phi) \equiv \frac{{M_p}^2}{2}\left(\frac{V_\phi}{V}\right)^2 \ll 1, \quad \eta(\phi) \equiv {M_p}^2\left(\frac{V_{\phi\phi}}{V}\right) \ll 1,
\end{equation}
where subscripts of $\phi$ denote derivatives with respect to the scalar field. 

For natural inflation, the parameters $\varepsilon$ and $\eta$ become
\begin{equation}
  \varepsilon(\phi) = \frac{1}{2}\left(\frac{M_p}{f}\right)^2\left[\frac{\sin(\phi/f)}{1-\cos(\phi/f)}\right]^2
\end{equation}
and
\begin{equation}
  \eta(\phi) = \left(\frac{M_p}{f}\right)^2\left[\frac{\cos(\phi/f)}{1-\cos(\phi/f)}\right].
\end{equation}
To first order, the spectral index $n_s$ and the tensor-to-scalar ratio $r$ can then be calculated using (see e.g. Ref.~\cite{Liddle:2000}),
\begin{equation}
\label{ns}
  n_s = 1-6\varepsilon+2\eta
\end{equation}
and
\begin{equation}
  r = 16\varepsilon.
\end{equation}
The predicted $(n_s, r)$ for natural inflation is consistent with the Planck data for $f \gtrsim 5 M_p$~\cite{Ade:2013rta}. 

\subsection{Multi-natural inflation}
In multi-natural inflation we consider an inflaton potential that consists of two or more sinusoidal functions. 
As a minimal extension, let us consider a potential with two sinusoidal terms.
The potential takes 
the form\footnote{
The potential of this form with $f_1/f_2$ being an irrational number was considered in Ref.~\cite{Banks:1991mb} in a context of
solving the cosmological constant problem.
}
\begin{equation}
  V(\f) = C - \Lambda_1^4\cos(\f/f_1) - \Lambda_2^4\cos(\f/f_2 + \theta),
  \label{eq:eni}
\end{equation}
where $C$ is a constant that shifts the minimum of the potential to zero and $\theta$ is a non-zero relative phase. 
The last term shifts the potential minimum from the origin to $\phi = \phi_{\rm min}$, and also modifies the potential shape.
This model is reduced to the original natural inflation in the limit of either $\Lambda_2 \rightarrow 0$ or $f_2 \rightarrow \infty$.
As we shall see later, such two sinusoidal terms could be generated by two different non-perturbative sources.
To simplify notation we set $f_1 = f$ and $\Lambda_1 = \Lambda$, and
relate the parameters by,
\begin{align}
f_2 &= A f,\\
\Lambda_2^4 &= B \Lambda^4,
\end{align}
where $A$ and $B$ are real and positive constants. Although in general $\Lambda_2$, $f_2$ and $\theta$ are arbitrary parameters, 
we only investigate cases for which the second sinusoidal term gives relatively small perturbations, and
the resulting potential is free of local minima so as to avoid the inflaton becoming trapped in a false
vacuum.
In general there are many other local minima and maxima at different values of $\phi$. The stability of the vacuum at $\phi = \phi_{\rm min}$ is assumed in the following analysis.

We have solved the inflaton dynamics numerically. To be explicit, we have solved the inflaton equation of motion,
$\ddot{\phi}+3H {\dot{\phi}} + V'(\phi) = 0$,  with the inflaton potential given by (\ref{eq:eni}), until the end of the
accelerated expansion of the Universe. Then, we have identified the timing when the number of e-folds until the
end of the inflation $N$ is equal to $50$ or $60$, i.e.,  when the cosmological perturbations with the pivot scale, $k = 0.002$\,Mpc$^{-1}$,
exited the horizon, and evaluated the slow-roll parameters at that time. We thus obtain the 
predicted values of $(n_s, r)$ for a given inflaton potential. By repeating this procedure for different values of $f$, we obtain a
line in the $(n_s, r)$-plane. 

\begin{figure}[t]
\begin{center}
\includegraphics[scale=0.382]{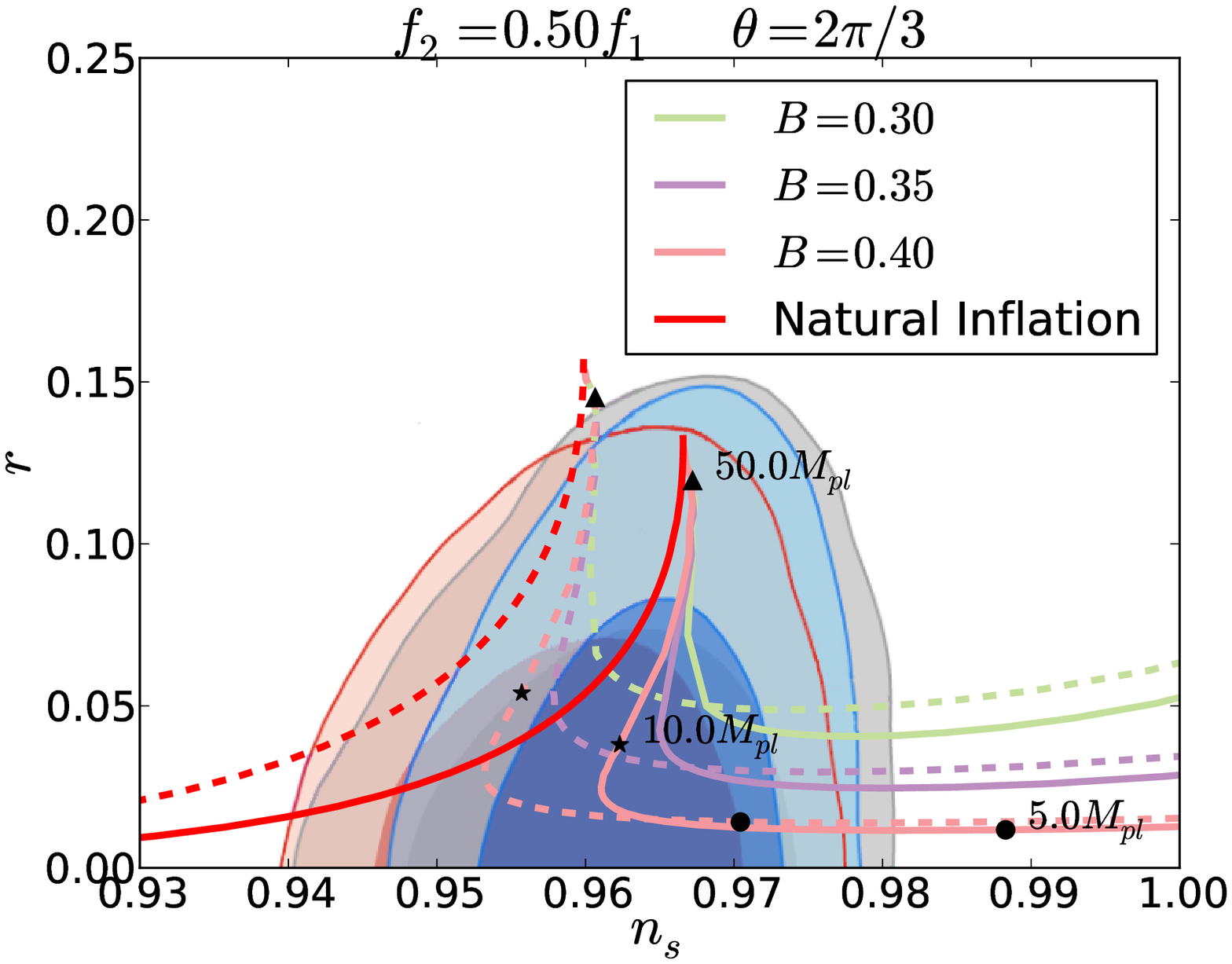}
\includegraphics[scale=0.382]{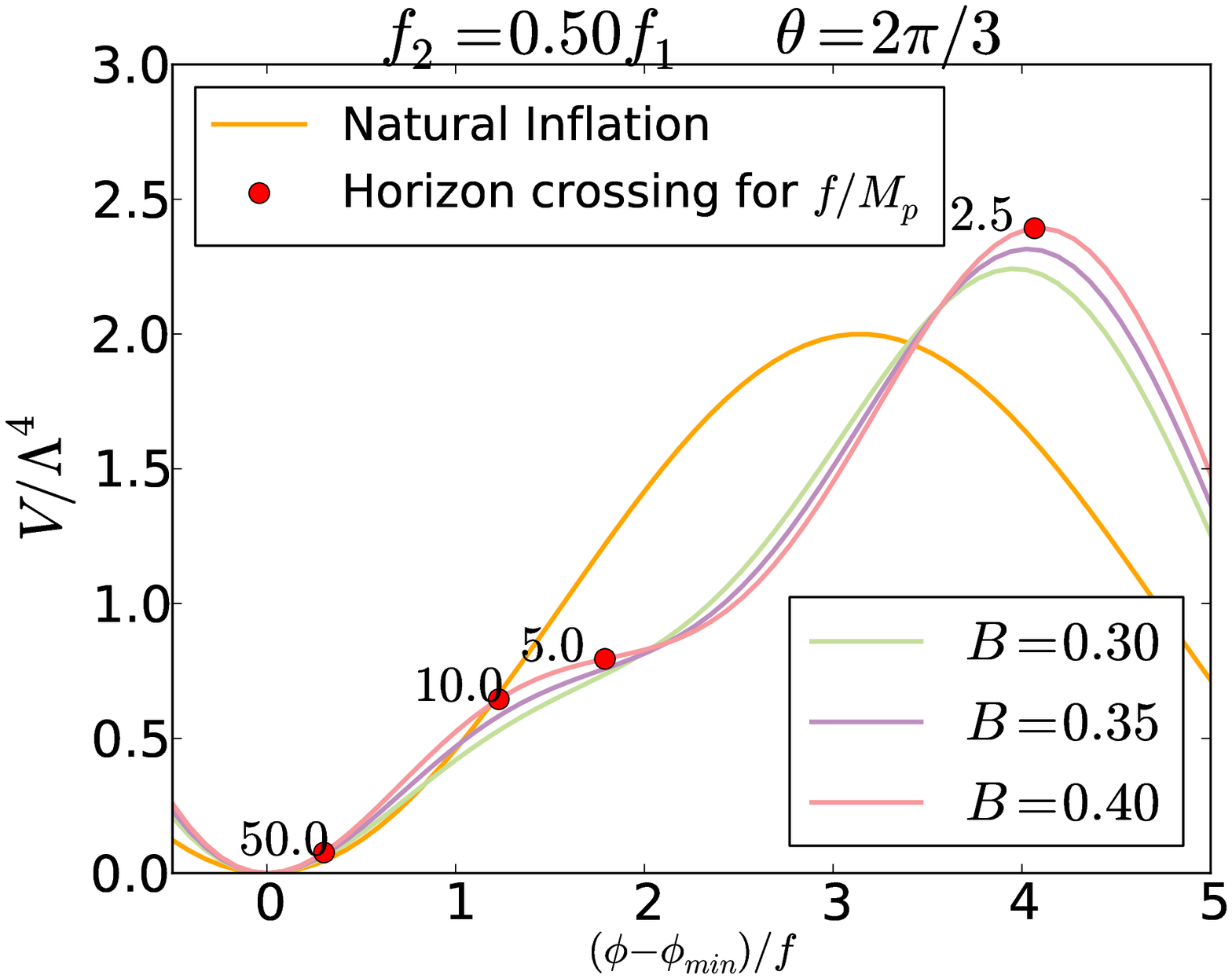}
\caption { 
Left: the prediction of $(n_s, r)$ of multi-natural inflation for three different values of $\Lambda_2^4$. 
Solid (dashed) lines correspond to the e-folding number $N=60$ ($N=50$). Right: the corresponding inflaton potentials. The red dots
represent the position at horizon crossing at $N= 60$ for the case of $B=0.40$.
}
\label{fig:nsr1}
\end{center}
\end{figure}

\begin{figure}[t]
\begin{center}
\includegraphics[scale=0.38]{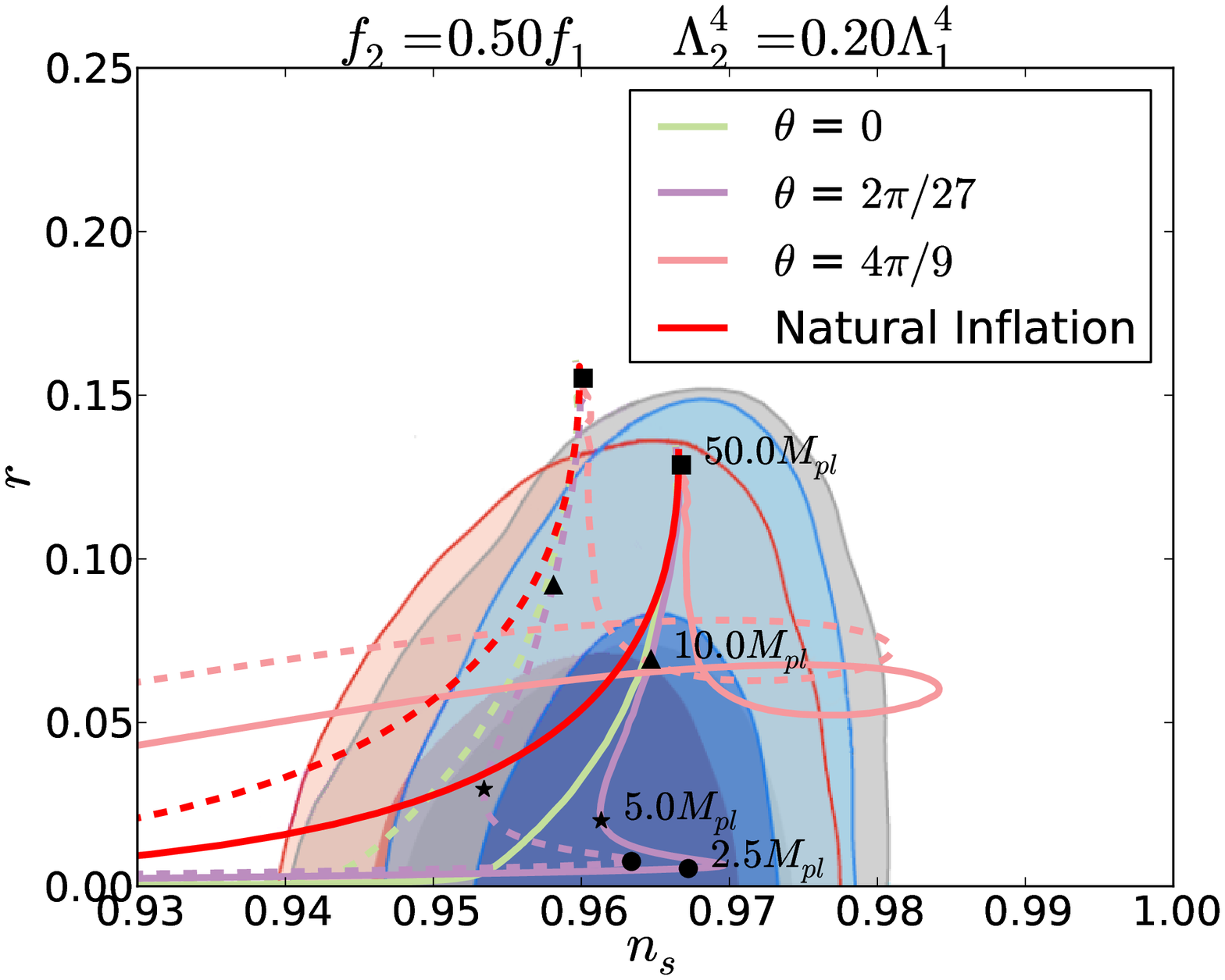}
\includegraphics[scale=0.38]{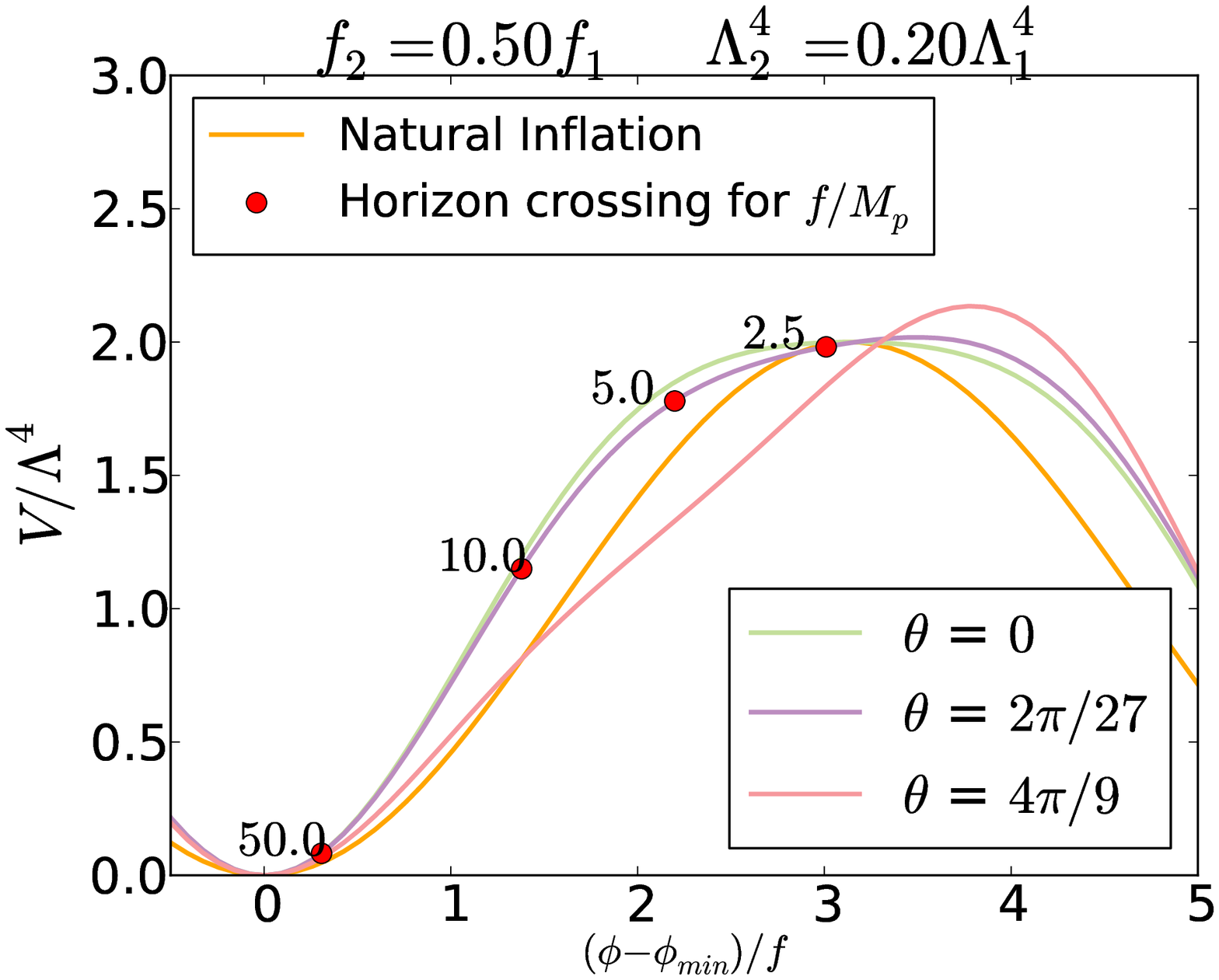}
\caption { 
Same as Fig.~\ref{fig:nsr1} but for different values of the relative phase $\theta$.
}
\label{fig:nsr2}
\end{center}
\end{figure}

The resulting $n_s$ and $r$ predictions  for varying values of $\Lambda_2$ and $\theta$ are 
shown in Figs.~\ref{fig:nsr1} and \ref{fig:nsr2}, respectively. Their corresponding potentials along with the positions on the potential for e-folding number $N=60$ for various values of $f/M_p$ are also shown.

In Fig.~\ref{fig:nsr1}, we set $A = 0.50$ and $\theta = 2\pi/3$, varying $B = 0.30, \,0.35$ and $0.40$.
From the left panel, we can see that the predicted $(n_s, r)$ approach those of the original natural inflation as $f$ increases.
This is because, in the limit of large $f$, both models are reduced to the quadratic chaotic inflation. Interestingly, for moderately large
$f$, the predicted curves  come closer to the center values of the Planck results, compared to the natural inflation.
We also note that, for $f \gtrsim 5 M_p$,  the behavior  of $(n_s, r)$ is similar to that of the polynomial chaotic inflation~\cite{Nakayama:2013jka}. 
This is not surprising because, if one expands the inflaton potential around the potential minimum,  multi-natural inflation 
can be approximated by the polynomial chaotic inflation for some choice of the model parameters. 
 In the right panel, we see that, for smaller values of $f$, the perturbation crosses the horizon scale further up the potential and, consequently, where $V_\phi/V$ is relatively small. Since $r = 16\varepsilon \propto (V_\phi/V)^2$, $r$ decreases as $f$ decreases. The behavior of $n_s$ is more complicated since both the slope and curvature of the potential (through $\eta \propto V_{\phi \phi}/V$) play a role. As $f$ becomes very large ($f \gtrsim 10M_p$), the potential at horizon crossing is close to the minimum and thus approaches the standard natural inflation solution. For $5 M_p \lesssim f \lesssim 10M_p$, the $\epsilon$-term in $n_s$ (see Eq.~(\ref{ns})) becomes less important 
and $n_s$ increases. As $f$ decreases below $\sim 5M_p$, $\varepsilon$ again begins to grow and thus so does $r$. Simultaneously, this causes $n_s$ to decrease as $\eta$ becomes negative. This causes the ($n_s, r$) curve to loop back toward negative $n_s$, similar to the pink curve in Fig.~\ref{fig:nsr2}. In this regime, however, the predicted $(n_s,r)$ is well outside the allowed region and thus we neglect it in Fig.~\ref{fig:nsr1}.

Next let us consider another case in which we set $A = 0.50$ and $B = 0.20$, varying the relative phase $\theta$ as $\theta = 0,\, 2\pi/27$, and
$ 4\pi/9$.  The behavior of $n_s$ and $r$ can be understood similarly. Compared to the previous case, there is a flat plateau 
around $\phi-\phi_{\rm min} \simeq \pi f$. This feature keeps the contribution of $\eta$ to the spectral index small even for small values of $f$.
As a result, the model is consistent with the Planck data for smaller values of $f$. In particular, 
the model agrees with Planck data in the $2\sigma$ region even for sub-Planckian values of $f$.

\begin{figure}[t]
\begin{center}
\includegraphics[scale=0.38]{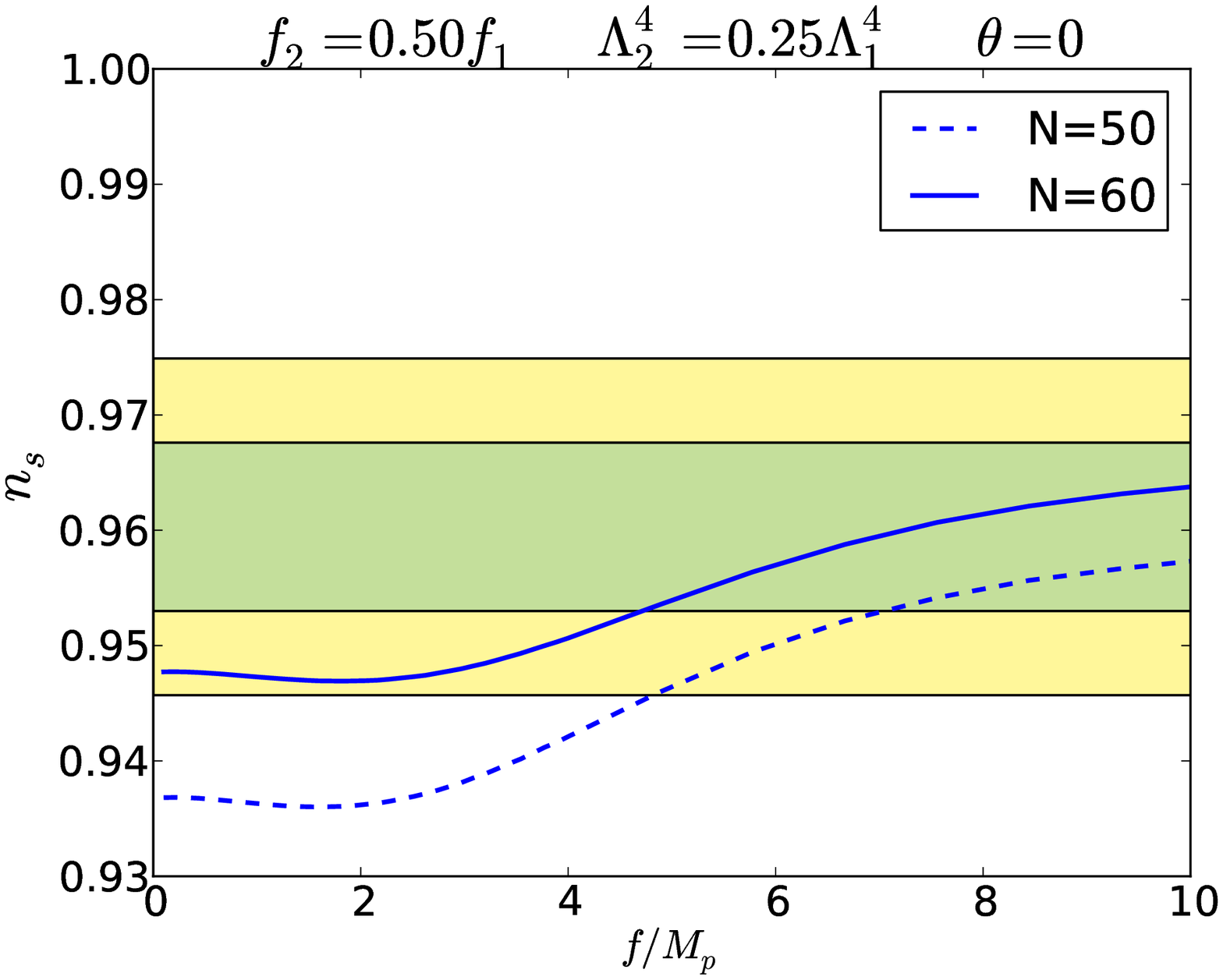}
\includegraphics[scale=0.38]{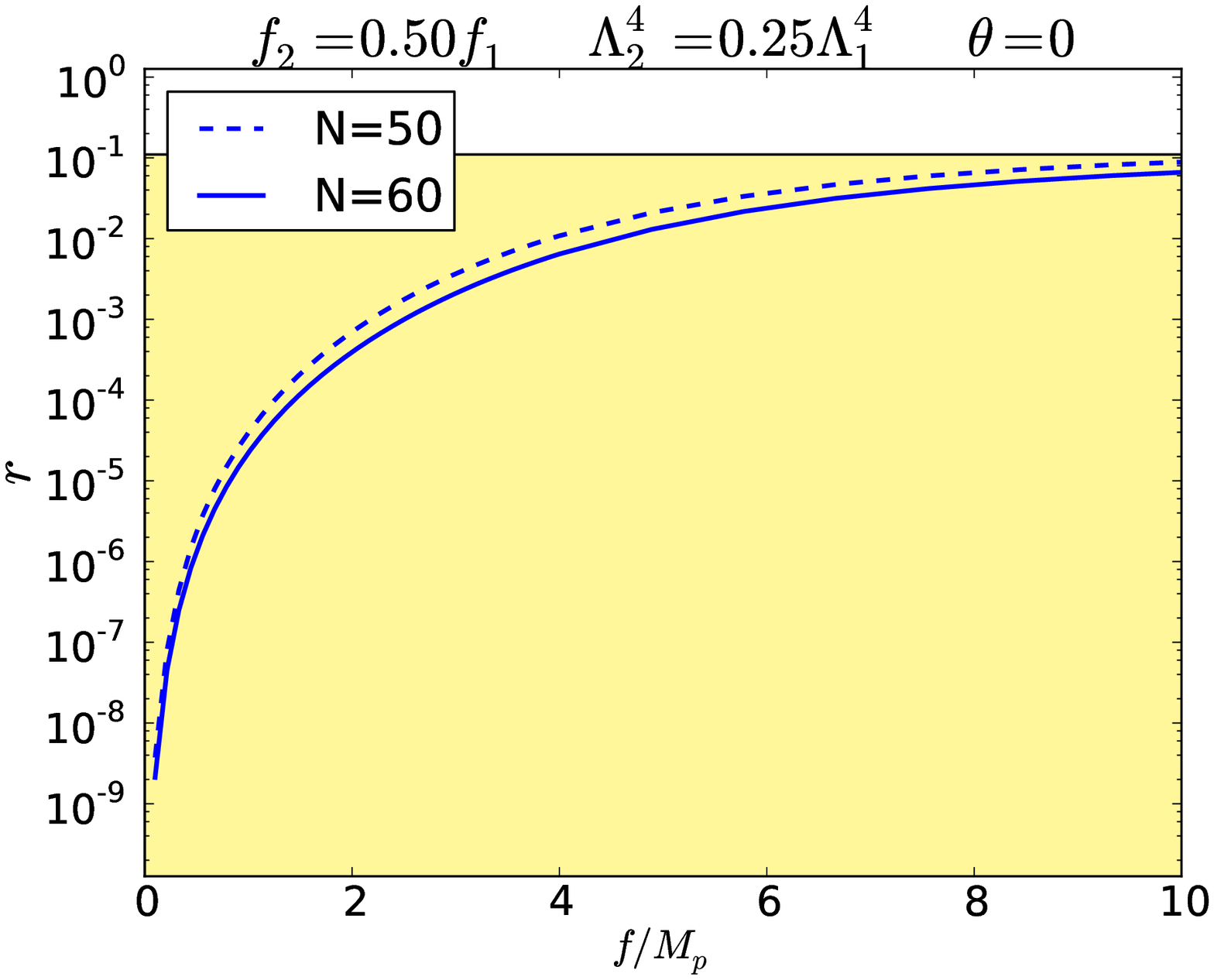}
\caption { 
Behavior of $n_s$ (left) and $r$ (right) as a function of $f$. The shaded regions in the left figure correspond to 1 and 2$\sigma$ allowed regions for $n_s$ from Planck data. The shaded region on the right corresponds to the 95\% CL for $r$ ($r < 0.11$).
}
\label{fig:nsrf}
\end{center}
\end{figure}

Our model (\ref{eq:eni}) actually contains a hilltop quartic inflation
model~\cite{Ade:2013rta, Linde:1981mu}   in a limiting case. The model is also referred to as new inflation. To see this,
let us investigate potentials where the curvature vanishes at the top of the first sinusoidal term, $\phi/f = \pi$. The condition for vanishing curvature imposes the conditions
\beq
B = A^2, \quad  \theta = - \frac{\pi}{A}.
\label{eq:cond}
\eeq
Expanding (\ref{eq:eni}) around $\phi/f = \pi$ our potential takes the form,
\begin{align}
V 
 &= \Lambda'^4 \left[1 - \frac{{\tilde \phi}^4}{\mu'^4} + \ldots \right]
\end{align}
where we have defined ${\tilde \phi} \equiv \phi - \pi f$ and
\begin{align}
\Lambda'^4 & \equiv C + \Lambda^4(1-A^2),\\
\frac{\Lambda'^4}{\mu'^4} & \equiv  \frac{(1-A^2) \Lambda^4}{24 A^2 f^4}.
\end{align}
For small $f \lesssim M_p$,  cosmological scales exit the horizon while the inflaton sits near the top of the potential
where the above expansion is valid.
Interestingly, this approximation is similar to hilltop models of inflation, given by:
\beq
V \simeq \Lambda^4 \left[1 - \frac{\phi^p}{\mu^p} + \ldots \right] 
\eeq
where $p = 4$ in our approximation. For $p = 4$, the spectral index $n_s$ is given by,
\beq
n_s \simeq 1- \frac{3}{N}
\eeq
which for $N=50$ ($N=60$) is 0.94 (0.95).
In Fig.~\ref{fig:nsrf} we show the behavior of $n_s$ and $r$ as a function of $f$
for the parameters $A = 0.50$, $B = 0.25$ and $\theta = 0$. 
We see that, indeed, for the given parameter set and sub-Planckian values of $f$, 
multi-natural inflation approaches the  hilltop quartic prediction.  For $N=60$, 
no lower bound on $f$ is required by the Planck data. For $N=50$, however, 
$f$ is required to be larger than $5 M_p$ as  it predicts too small a value of $n_s$. 
As the e-folding number becomes smaller for smaller $f$, 
this tension at small values of $f$ is real and requires some modification of the inflaton dynamics. 
In a context of new inflation in supergravity~\cite{Kumekawa:1994gx,Izawa:1996dv}, the resolution
of the tension was discussed in detail, and it is known that the prediction of $n_s$ can be increased to be consistent with the Planck data either by
adding a logarithmic correction~\cite{Nakayama:2011ri,Bose:2013kya} or a linear term~\cite{Takahashi:2013cxa}, or by
considering $p > 4$~\cite{Harigaya:2013pla}. In the multi-natural inflation, we can add another sinusoidal function so as to
either give an effective linear term, or cancel both the curvature and the quartic coupling at the top of the potential.

\begin{figure}[t]
\begin{center}
\includegraphics[scale=0.37]{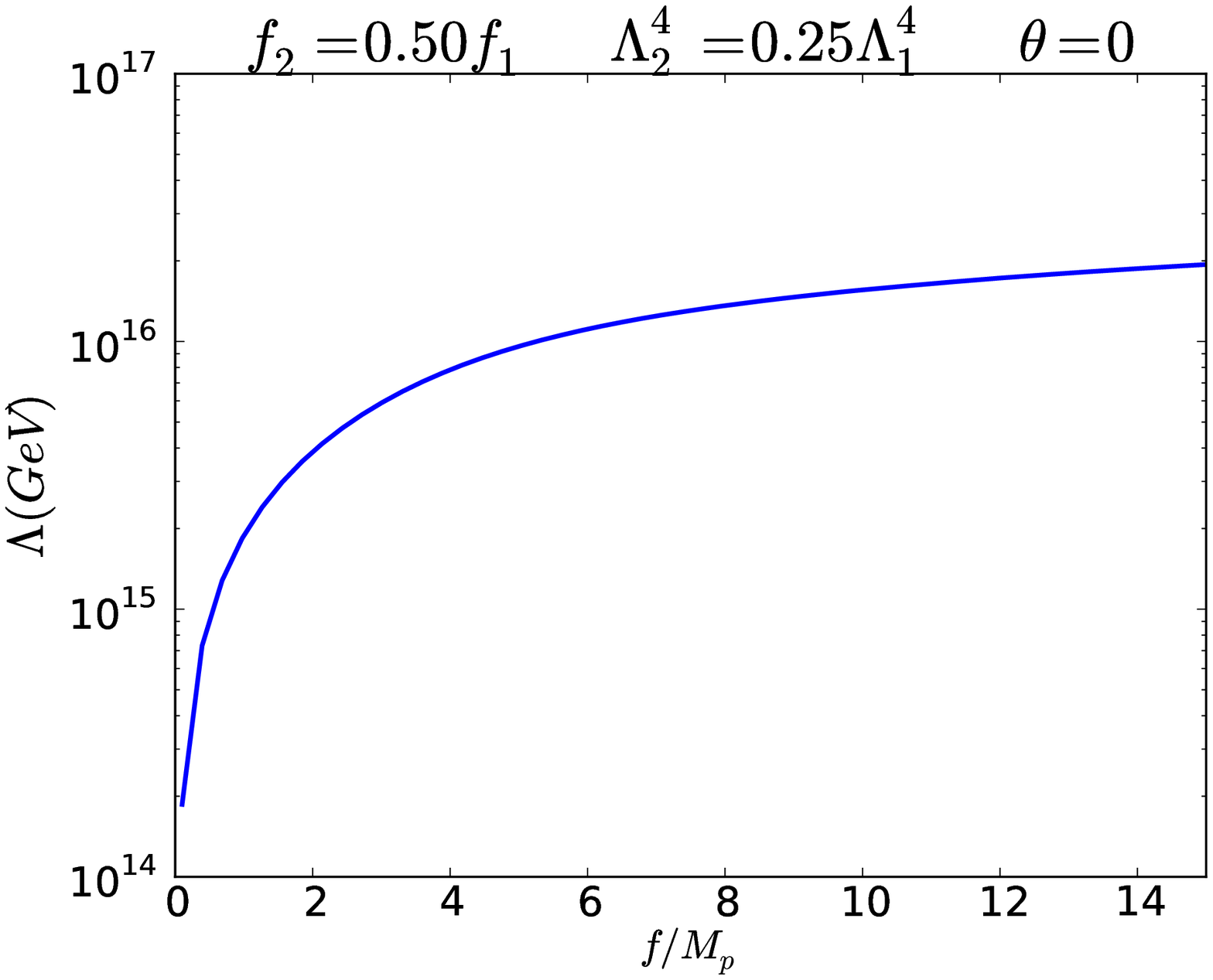}
\includegraphics[scale=0.37]{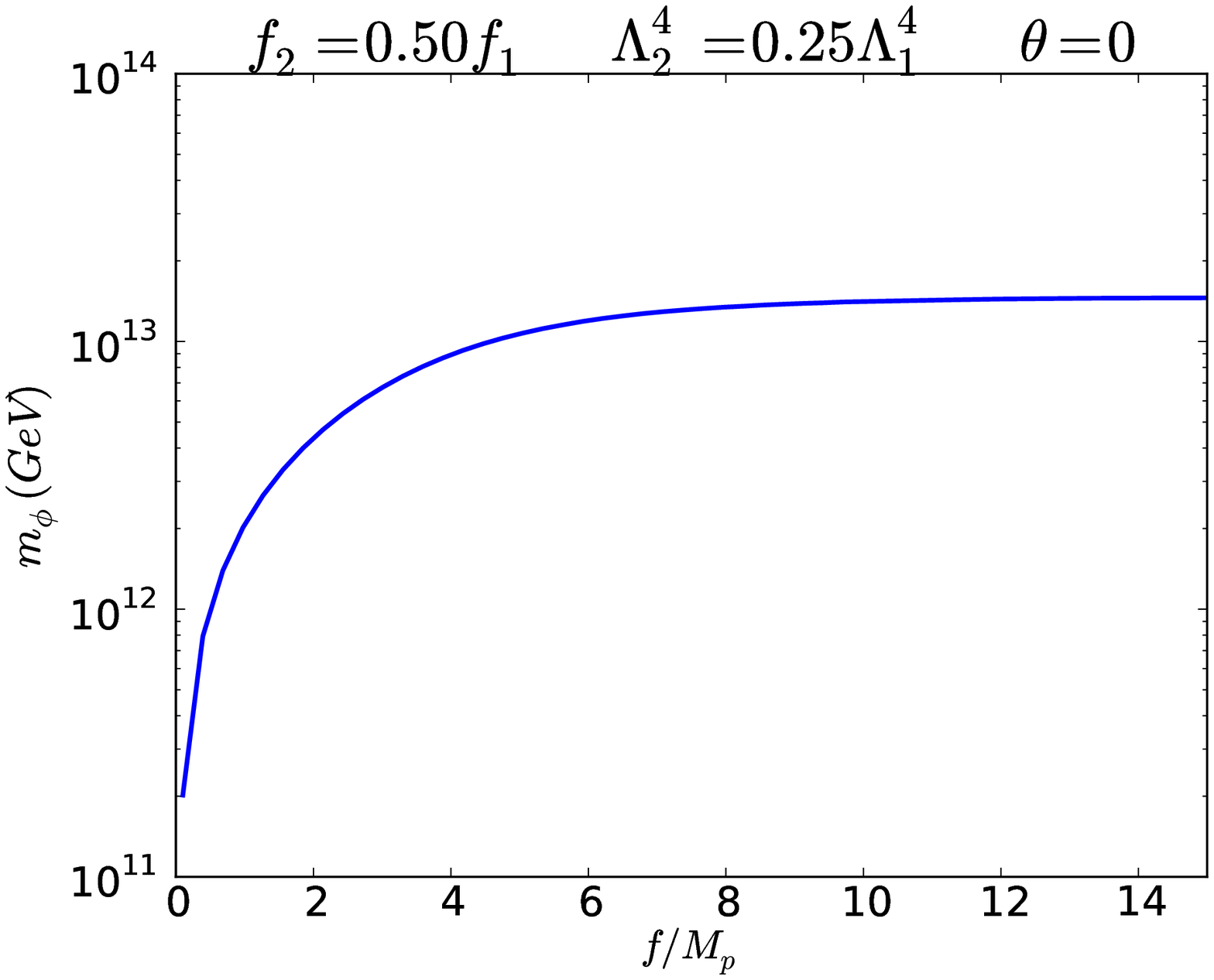}
\caption { 
Planck normalized values for $\Lambda$ (left) and $m_\phi$ (right) as a function of $f$.
}
\label{fig:COBE}
\end{center}
\end{figure}

We can constrain the height of the potential, ${\Lambda}^4 = \Lambda_1^4$, by imposing Planck normalization on the
curvature perturbation~\cite{Ade:2013rta}
\beq
P_{\mathcal R} \simeq e^{3.098} \times 10^{-10} \simeq 2.2 \times 10^{-9}.
\eeq
Following Ref.~\cite{Liddle:2000}, the scalar perturbation
amplitude is given by
\begin{equation}
   P_\mathcal{R}^{1/2} = \frac{H^2}{2\pi \left|\dot{\phi}\right|},
\end{equation}
where the right-hand side is to be evaluated when the comoving wavelength of the perturbation crosses outside the horizon during inflation. Using the
slow roll approximation, we find
\begin{equation}
   P_\mathcal{R}^{1/2} \simeq \frac{3}{2\pi}\left(\frac{1}{3}\right)^{3/2}\frac{\Lambda^2 f}{M_p^3}
   \frac{\left[C-\cos(\phi/f)-B\cos(\phi/(Af)+\theta)\right]^{3/2}}{\left|\sin(\phi/f)+(B/A)\sin(\phi/(Af)+\theta)\right|}.
\end{equation}
Fixing the parameters $A, B$ and $\theta$ we can find $\Lambda$ for varying $f$. As an example, Fig.~\ref{fig:COBE} shows the behavior of $\Lambda$ and the inflaton mass $m_\phi$ at the potential minimum for $0.5M_p \leq f \leq 15M_p$, again using the parameters $A = 0.50$, $B = 0.25$ and a zero relative phase. For large $f$ we see that $\Lambda \sim 10^{16}$ GeV and $m_\phi \sim 10^{13} $ GeV.\footnote{
We note that the inflaton mass at the potential minimum vanishes for $A^{-1} = 2 i +1$ with $i \in {\bf N}$.
This can affect the reheating process of the inflaton. 
}

\section{Discussion and Conclusions}
\label{sec:3}
We have extended natural inflation by adding another sinusoidal function.
The origin of such shift-symmetry breaking terms could be due to non-perturbative
effects. For instance, let us consider a complex scalar field $\Phi$ coupled to
two sets of quark and anti-quark fields,
\beq
\label{qQ}
{\cal L}\;=\; \sum_{i=1}^{n_q} y_i \Phi q_i {\bar q}_i + \sum_{j=1}^{n_Q} Y_j \Phi Q_j {\bar Q}_j, 
\eeq
where $y_i$ and $Y_j$ are coupling constants. Here 
$q_i$ and ${\bar q}_i$ are charged under a hidden non-Abelian gauge 
group $G_1$, while  $Q_j$ and ${\bar Q}_j$ under another non-Abelian group $G_2$. 
To be concrete we take $G_1 = $SU$(N_1)$ and $G_2 = $SU$(N_2)$, and those (anti-)quarks
are in the (anti)fundamental representation of each gauge group.
Let us suppose that $\Phi$ respects a global U(1) symmetry, $\Phi \rightarrow \Phi e^{i \alpha}$,
which is explicitly broken by non-perturbative effects of the two gauge interactions.
If $\Phi$ develops a non-zero vacuum expectation value (vev), its phase component becomes a 
Nambu-Goldstone boson, which we identify with the inflaton $\phi$:
\beq
\Phi = \frac{v + {\hat s}}{\sqrt{2}} \exp\left[i\frac{\phi}{v}\right],
\eeq
where $v$ is the vev of $\Phi$, and ${\hat s}$ is the radial component. If ${\hat s}$ is heavy enough, 
it can be integrated out. In the low energy limit, both $G_1$ and $G_2$ become strong, producing a 
potential for $\phi$:
\beq
V(\phi) \;=\; \Lambda_1^4 \cos\left(\frac{\phi}{f_1}\right)+\Lambda_2^4 \cos\left(\frac{\phi}{f_2}\ + \theta \right),
\eeq
where $\Lambda_1$ and $\Lambda_2$ denote the dynamical scale of $G_1$ and $G_2$, respectively. 
The decay constants $f_1$ and $f_2$ are related to the vev of $\Phi$ as
\beq
f_1 = \frac{v}{n_q},~~~~f_2 = \frac{v}{n_Q}.
\eeq
Thus, our model (\ref{eq:eni}) corresponds to $A = n_q/n_Q$ and $B=\Lambda_2^4/\Lambda_1^4$.
Note that $A$ is  a rational number in this case. In order to satisfy the Planck normalization, the dynamical scale
should be of order $\GEV{15-16}$ for $f_1 \sim f_2 \gtrsim {\cal O}(0.1)M_p$. 
Such a field theoretic description is considered to be valid for the decay constant $f$ smaller than the Planck scale. 
As we have seen in the previous section, sub-Planckian values of $f$ can fit the Planck data. 
Therefore, multi-natural inflation can be easily realized in an effective field theory. 

It may be possible to embed multi-natural inflation into string theory, where an axionic
component of moduli fields is identified with the inflaton, since are various sources for the inflation 
potential such as gaugino condensations and instantons. This issue will be discussed in details
elsewhere~\cite{Czerny:2014xja}.

We have so far focused on the case of two sinusoidal functions (\ref{eq:eni}). It is straightforward to consider multiple 
sinusoidal functions. 
In order to have successful inflation for the decay constants of order the Planck scale or below, we need to arrange those sinusoidal
functions so that they conspire to make the inflaton potential sufficiently flat.
As one of the slow-roll parameter $\eta$ is of order unity in the original natural 
inflation with $f \sim M_p$,  this requires a typical fine-tuning of ${\cal O}(1)$\%.
As we have seen in the previous section, considering more than two sinusoidal functions, we can increase the prediction
of $n_s$ to be consistent with the Planck data for small values of $f$. 

The inflaton needs to have couplings with the standard model particles for successful reheating. We may introduce dilatonic couplings,
\beq
{\cal L} = c \frac{\phi}{f}\, F_{\mu \nu} {\tilde F}^{\mu \nu},
\eea
where $c$ is a coupling constant, and $F_{\mu \nu}$ denotes the field strength of the standard-model gauge fields. Such interactions can be induced if the heavy
quarks $q$ or $Q$ in \EQ{qQ} are also charged under the standard-model gauge symmetry. For $f \sim M_p$ and $m_\phi \sim \GEV{13}$,
the resultant reheating temperature is expected to be of order $\GEV{8-10}$, depending on the size of $c$. The thermal leptogenesis~\cite{Fukugita:1986hr} 
is possible for such high reheating temperature.

In this letter we have studied a multi-natural inflation model where the single-field inflaton 
potential is given by
a sum of two or more sinusoidal potentials comparable in size but with
a slightly different periodicity as well as a possible non-zero relative phase. 
The model is versatile enough to realize both large-field and
small-field inflation.
We have shown that, in a model with two sinusoidal terms, 
the predicted values of the spectral index and the tensor-to-scalar ratio are
consistent with the Planck data; they come closer to the center values of the Planck results,
compared to natural inflation. The on-going and planned B-mode polarization experiments will be 
able to probe a large portion of the parameter space of our model. 
We have also shown that our model can be approximated by the  hilltop quartic model in a limiting case. In particular, 
the spectral index lies within the $2\sigma$ allowed region even for decay constants much smaller than 
the Planck scale. Therefore there is no lower bound on the decay constants. 
 This result should be contrasted to the original natural inflation, which is consistent with the Planck data only for
decay constants greater than $\sim 5 M_p$. 
This eases the difficulty of implementing the multi-natural inflation in an effective field theory.

\vspace{5mm}
{\it Note added:}
After the submission of our letter, the BICEP2 experiment found the primordial B-mode
polarization~\cite{Ade:2014xna}. Although it needs to be confirmed by other experiments,
this discovery makes it plausible that our model can be supported or refuted by future
observations.

\section*{Acknowledgments}
FT thanks Tetsutaro Higaki for discussion.
This work was supported by Grant-in-Aid for  Scientific Research on Innovative
Areas (No.24111702, No. 21111006, and No.23104008), Scientific Research (A)
(No. 22244030 and No.21244033), and JSPS Grant-in-Aid for Young Scientists (B)
(No. 24740135), and Inoue Foundation for Science.
This work was also supported by World Premier International Center Initiative
(WPI Program), MEXT, Japan [FT].



\end{document}